\documentclass[a4paper]{article}
\usepackage{epsfig}
\title{ Investigation of the Spin-Peierls transition in 
CuGeO$_3$ by Raman scattering }

\author{P. Lemmens, B. Eisener, M. Brinkmann, L.V. Gasparov,\\
G. G\"untherodt, P.v. Dongen$^+$, W. Richter$^*$,\\ 
M. Weiden$^*$,C. Geibel$^*$, F. Steglich$^*$}

\begin{document}
\maketitle

\centerline{2. Physikalisches Institut, RWTH Aachen, }
\centerline{$^+$Institut f\"ur Theoretische Physik, RWTH Aachen,
52056 Aachen, Germany} 
\centerline{$^*$Institut f\"ur Technische Physik, TH Darmstadt,
64289 Darmstadt, Germany}

Proceedings of SCES'1995, Goa, India, to be published in Physica B (1996)

\begin{abstract}
Raman experiments on the spin-Peierls compound CuGeO$_3$ and 
the substituted (Cu$_{1- x}$,Zn$_x$)GeO$_3$ and Cu(Ge$_{1-x}$,Ga$_x$)O$_3$ 
compounds were performed in order to investigate the response of specific 
magnetic excitations of the one-dimensional spin-1/2 chain to spin 
anisotropies and substitution-induced disorder. In pure CuGeO$_3$, 
in addition to normal phonon scattering which is not affected at all by 
the spin-Peierls transition, four types of magnetic scattering features 
were observed. Below T$_{SP}$=14 K a singlet-triplet excitation at 
30 cm$^{-1}$, two-magnon scattering from 30 to 227 cm$^{-1}$ and folded 
phonon modes at 369 and 819 cm$^{-1}$ were identified. 
They were assigned by their temperature dependence and lineshape. 
For temperatures between the spin-Peierls transition T$_{SP}$ and 
approximately 100 K a broad intensity maximum centered at 300 cm$^{-1}$ 
is observed. The temperature dependence of this intensity is similar to 
the behavior of lattice fluctuations recently observed in an electron 
diffraction study. This scattering is attributed to dynamical 
spin-Peierls fluctuations of the weakly or non-static dimerized spin chain.

\end{abstract}

\footnote{

Author for correspondence: 

Dr. P. Lemmens, 2. Physikalisches Institut RWTH Aachen, 
52056 Aachen, Germany, \\Tel.: 0241-807113, Fax.: 0241-8888306
Email: lemmens@sun1.phys2.rwth-aachen.de}
\newpage
\section{Introduction }
It has been of continued interest to understand the physical properties 
of low-dimensional quantum spin systems to clarify the relation between 
dimensionality or anisotropies and low energy excitations. The recently 
discovered spin-Peierls (SP) transition in the compound CuGeO$_3$ [1], 
resulting from the coupling of an antiferromagnetic spin-1/2 Heisenberg 
chain to a three- dimensional lattice system, is a model system in this 
respect. The magnetic susceptibility of CuGeO$_3$ shows a broad maximum 
at about 60 K and an exponential, isotropic decrease below T$_{SP}$=14 K, 
reflecting the effect of quantum fluctuations above T$_{SP}$ and the 
continuous dimerization of the Heisenberg chain below T$_{SP}$ [2]. By 
inelastic neutron scattering the dispersion relations of magnetic
excitations, 
the anisotropy of the exchange coupling and the singlet-triplet transition 
$\Delta$(q$\approx$0)=16.8 cm$^{-1}$ of the dimerized Heisenberg chain 
were determined [3,4]. For temperatures above T$_{SP}$ specific heat 
measurements show that the magnetic entropy reaches R$\cdot$ln2 not until 
T$\approx$150 K is approached [5]. Additional hints toward strong lattice 
fluctuations along the spin-1/2 chains exist in electron diffraction 
experiments for temperatures T$\leq$100 K [6]. By Raman scattering, 
phonon and two-magnon scattering and the singlet-triplet excitation 
at 2$\Delta$=30 cm$^{-1}$ were identified [7, 8].

\section{Experimental} 
Polycrystalline samples of the stoichiometry (Cu$_{1-x}$,Zn$_x$)GeO$_3$ 
and \newline Cu(Ge$_{1- x}$,Ga$_x$)O$_3$ with x ranging from 0 to 0.07 and
0.04, 
respectively, were prepared by standard solid state reaction. Single 
crystals of CuGeO$_3$ were grown by slow cooling of a stoichiometric 
mixture. The samples were characterized by magnetic susceptibility 
measurements and x-ray diffraction. The magnetic susceptibility shows 
at the lowest temperatures measured ($\approx$ 2K) a small upturn, pointing to 
a negligible content of paramagnetic impurtities, which may result from 
disrupted Cu-chains.

\section{Results and Discussion}
In Fig.1 the scattering intensity of single crystalline CuGeO$_3$ 
is shown in (z,z)-geometry, i.e. with polarization of the incident 
and scattered light both parallel to the Cu-chains (c-axis of the 
crystal lattice). Below T$_{SP}$=14 K there appears additional 
scattering intensity  at 30, 105, 227, 369 and 819 cm$^{-1}$. The 
peak positions and widths of the modes at 30, 227 and 369 cm$^{-1}$ 
are given in Fig. 2. Comparing the scattering intensity with a 
two-magnon density of states calculated in Heisenberg spin-wave 
approximation we get reasonable agreement with the measured intensity 
in the interval 30-300 cm$^{-1}$. As fit parameters we used the 
following exchange coupling constants J$_c$=68.6 cm$^{-1}$, 
J$_b$=6.1 cm$^{-1}$, J$_a$=-0.5 cm$^{-1}$ and J'=10.4 cm$^{-1}$. 
The coupling constants J$_c$, J$_b$, J$_a$ along the principal 
axes of the compound and J' (giving the alternating exchange along 
the Cu-chain with J$_{1/2}$=$\pi$/2$\cdot$J$_c$$\pm$J') show 
quantitative agreement with neutron scattering and susceptibility 
measurements (J$_c$$^{susc}$=61 cm$^{-1}$) [1,3]. However, the 
sharp peak at 30 cm$^{-1}$ does neither fit in intensity nor in 
its temperature dependence into the concept of spin waves.
\footnote{Note added after proof: This intensity and the continuum
at $\approx$ 300 cm$^{-1}$ was
modeled recently by calculating frustration induced Raman scattering
of a one-dimensional
Heisenberg chain, see: V.N. Muthukumar, et al., Frustration induced
Raman scattering in CuGeO$_3$.}
This is shown in Fig. 2 by comparing its temperature dependence and 
width with the two-magnon scattering at 227 cm$^{-1}$ and the 
dimerization-induced folded phonon mode at 369 cm$^{-1}$. The 
singlet-triplet excitation shows a drastic frequency shift and 
linewidth broadening with increasing temperature while the other 
modes disappear without any renormalizations in frequency or 
linewidth. This large broadening comparable with the self-energy 
of the excitation is expected for magnetic excitons if their degree 
of localization depends strongly on the order parameter, i.e. the 
alternation of the exchange coupling [9]. In the noninteracting-dimer 
limit excitons with a bound character are well defined excitations. 
However, due to the weak exchange coupling and the inherent instability 
of the spin-chain a more dynamical pairing description of the triplet 
states comparable with a spin-Peierls t-J-model is suggested leading 
to weak or non-static dimerization [10,11]. A critical test of this 
state would be the observation of spin and lattice fluctuations of 
the Heisenberg chain above T$_{SP}$ with characteristic energies 
higher than the exchange coupling. Electron diffraction and specific 
heat measurements both show fluctuation contributions for temperatures 
T$\leq$150 K [5,6]. Further hints for unusual spin dynamics above 
T$_{SP}$ are resolved by Cu NQR and interpreted in a similar way [12]. 
Here we present evidence for these fluctuations in form of a broad 
maximum centered near 300 cm$^{-1}$ with a total width of 300 cm$^{- 1}$, 
as shown by the inset of Fig. 3. The energy corresponding to this maximum 
is a factor of 3-4 higher than the exchange coupling in the dimerized 
state and may therefore not be interpreted as two-magnon scattering. 
Its scattering contribution vanishes rapidly in the dimerized state 
leading to a redistribution of intensity to lower energy for 
T$\leq$T$_{SP}$. At higher temperatures, T$\geq$100 K the maximum 
is smeared out to higher and lower frequencies. In Fig. 3 we show 
the temperature dependence of this intensity, which was determined 
by subtracting the intensity at 520 cm$^{-1}$ from the intensity 
at the maximum (determined at 256 cm$^{-1}$). This subtraction 
excludes phonon scattering and small changes of the laser power. 
The intensity of the assigned fluctuation contributions in Fig.3 
shows a maximum at T$_{SP}$ and a sharp drop toward lower 
temperatures.

Further evidence for the non-static evolution of the spin-Peierls 
phase may result from the strong sensitivity of T$_{SP}$ to any 
substitution in the Cu-chains [13]. Our Raman experiments show that 
the singlet-triplet excitation at 2$\Delta$=30 cm$^{-1}$ exhibits 
the strongest sensitivity to substitutions. For Ga-contents larger 
than x=0.03 and Zn-contents larger than x=0.01 this scattering 
intensity vanishes completely while the folded phonon modes still 
survive. The latter intensity vanishes if the spin-Peierls state 
is completely destroyed by substitution. For smaller substitution 
levels a reduction of the energy of the singlet-triplet excitation 
up to 20\% is observed without any broadening of its lineshape. 
This is related to the pinning of spin fluctuations at the 
substitution site [8].

This work was supported by DFG through SFB 341, SFB 252 and 
by BMBF FKZ 13N6586
\newpage
\section{References} 

[1] M. Hase et al., Phys. Rev. Lett. 70 (1993) 3651 and 
Phys. Rev. B48, 13 (1993) 9616
\newline
[2] J.C. Bonner and M.E. Fisher, Phys. Rev. 135 (1964) A640\newline
[3] M. Nishi, O. Fujita, J. Akimitsu, Phys. Rev. B50, 9 (1994) 6508\newline 
[4] O. Fujita et al., Phys. Rev. Lett. 74, (1994) 1677\newline 
[5] M. Weiden, J. K\"ohler, G. Sparn, M. K\"oppen, M. Lang, C. Geibel, 
F. Steglich, to be published \newline
[6] C.H. Chen and C-W.Cheong, Phys. Rev. B51, 10 (1995) 6777\newline
[7] H. Kuroe et al., Phys. Rev B22, 50 (1994) 16468\newline
[8] P.Lemmens, B.Eisener, M.Brinkmann, L.V. Gasparov, G. G\"untherodt, 
P.v. Dongen, W.Richter, M.Weiden, C. Geibel, F.Steglich,
to be published\newline
[9] J.C. Bonner, H.W.J. Bl\"ote, Phys. Rev. B25, 11 (1982) 6959\newline
[10] M. Imada, J. Phys. Soc. Jpn. 60 (1991) 1877 and 61 (1992) 423\newline 
[11] P.W. Anderson, Mater. Res. Bull. 8 (1973) 153 \newline
[12] J. Kikuchi, et al.,  J. Phys. Soc. Japan, 63, 3 (1994) 872\newline
[13] S.B. Oseroff, et al., Phys. Rev. Lett. 74, 8 (1995) 1450\newline

\begin{figure}[b]
\centerline{\psfig{file=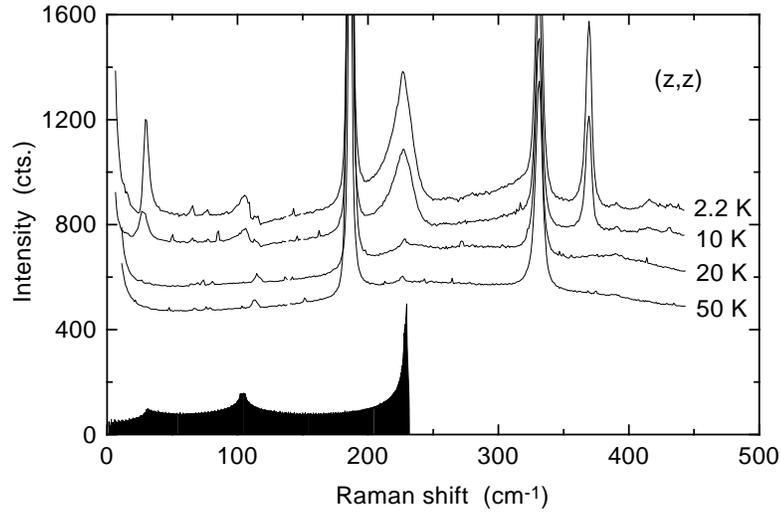,width=12.2cm}}
\caption{
Raman scattering intensities of single crystalline 
CuGeO$_3$ at 2.2, 10, 20 and 50 K. The curves are shifted by 
an overall offset of 400 counts and by 100 counts between the 
individual spectra. The dark area shows a calculation of the 
two-magnon density of states in the spin-Peierls phase with the 
exchange coupling constants J$_c$=68.6 cm$^{-1}$, 
J$_b$=6.1 cm$^{-1}$, J$_a$=-0.5 cm$^{-1}$ and J'=10.4 cm$^{-1}$.} 
\end{figure}

\begin{figure}
\centerline{\psfig{file=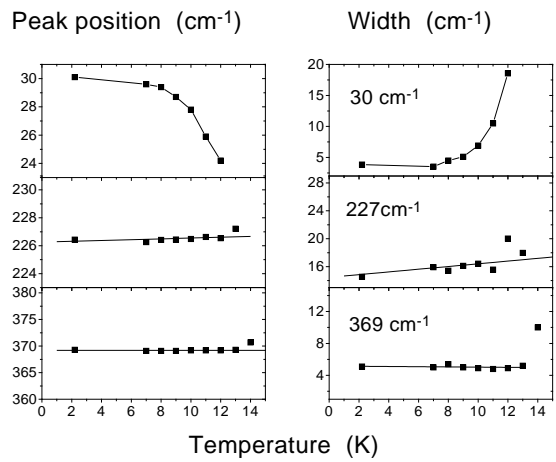,width=12.2cm}}
\caption{Peak position and width (FWHM) of the singlet-triplet 
excitation at 30 cm$^{-1}$, the two-magnon scattering at 227 
cm$^{-1}$ and a folded phonon mode at 369 cm$^{-1}$ as a 
function of temperature. The lines are guides to the eye.}
\end{figure}

\begin{figure}
\centerline{\psfig{file=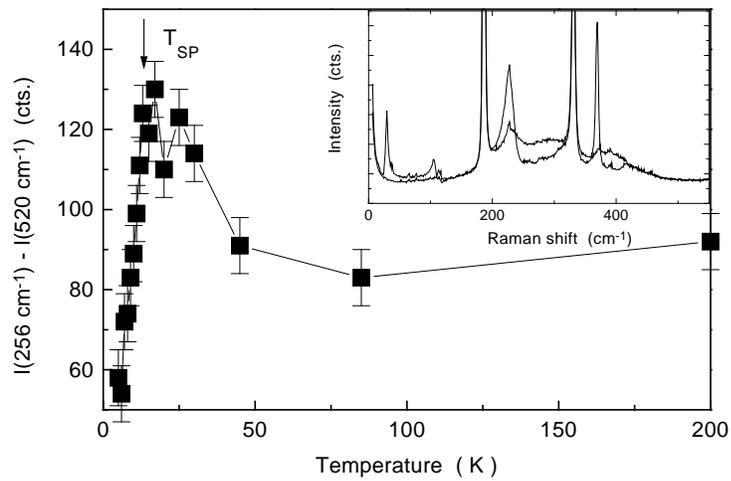,width=12.2cm}}
\caption{Scattering intensity of the broad maximum near 300 
cm$^{-1}$ as a function of temperature 
determined by subtracting the intensity at 520 cm$^{-1}$ from 
the intensity at 256 cm$^{-1}$. The line is a guide to the eye. 
The inset shows Raman spectra at 7 K (lower curve) and 15 K (upper curve).}
\end{figure}
\end{document}